\documentclass[aps,twocolumn,pra,superscriptaddress,letterpaper,nofootinbib]{revtex4-1}
\usepackage{amssymb}
\usepackage{mathtools}  
\usepackage[pdftex]{graphicx}
\usepackage[usenames, dvipsnames, svgnames, table]{xcolor}
\usepackage[colorlinks=true, citecolor=Blue,
            linkcolor=BrickRed, urlcolor=ForestGreen]{hyperref}
\usepackage{mathptmx}
\usepackage{bm}
\usepackage{natbib}

\begin{document}


\newcommand{\R}[1]{\textcolor{red}{#1}}
\newcommand{\B}[1]{\textcolor{blue}{#1}}
\renewcommand{\thefootnote}{\fnsymbol{footnote}}


\title{General quantum
constraints on detector noise in continuous linear measurement}

\author{Haixing Miao}
\affiliation{School of Physics and Astronomy,
University of Birmingham, Birmingham, B15 2TT, United Kingdom}


\begin{abstract}

In quantum sensing and metrology, an important class of
measurement is the continuous linear measurement, in which
the detector is coupled to the system of interest linearly and
continuously in time. One key aspect involved is the quantum noise of the
detector, arising from quantum fluctuations in the detector
input and output. It determines how fast we acquire
information about the system, and also influences the system evolution
in terms of measurement back action. We therefore often categorize it
as the so-called imprecision noise and quantum back action noise.
There is a general Heisenberg-like uncertainty relation
that constrains the magnitude of and the correlation between
these two types of quantum noise. The main result of
this paper is to show that when
the detector becomes ideal, i.e., at the quantum limit
with minimum uncertainty,
not only does the uncertainty relation takes the equal sign as
expected, but also there are two new equalities.
This general result is illustrated by using
the typical cavity QED setup with the system
being either a qubit or
a mechanical oscillator.
Particularly, the dispersive readout of
qubit state, and the measurement
of mechanical motional sideband asymmetry are considered.
\end{abstract}

\maketitle

\section{Introduction and summary}

When we probe classical signals or quantum systems,
noise in the detector limits our ability to extract the
relevant information. If the classical noise
is sufficiently suppressed, the detector will enter
the regime where intrinsic quantum fluctuations in its
degrees of freedom
determines the statistical property of the noise.
Modern experiments are approaching such
a quantum-noise-limited regime\,\cite{Clerk2008, Giovannetti2011}.
The state-of-the-art
includes, e.g., high-fidelity qubit readout\,\cite{Jeffrey2014},
gravitational-wave detection using laser
interferometers\,\cite{lrr-2012-5,
Adhikari2014}, and quantum
optomechanics in general\,\cite{Yanbei:Review, Aspelmeyer2014}.

\begin{figure}[!b]
\includegraphics[width=0.45\textwidth]{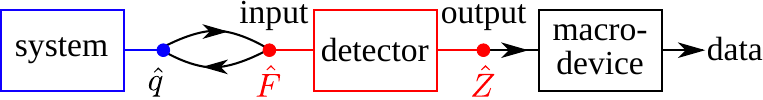}
\caption{A schematics for the continuous linear measurement.
\label{fig:measurement_model}}
\end{figure}

In the quantum regime,
those experiments mentioned above can be modeled as the
continuous linear quantum measurement, which is represented
schematically in Fig.\,\ref{fig:measurement_model}. The system can
be, e.g. either a qubit or a mechanical oscillator,
which can be further attached to a classical signal if
acting as a sensor. The detector is a quantum field that
contains many degrees of freedom, e.g. the optical field
in optomechanics, and interacts with the system continuously
in time. We call the degree of freedom linearly
coupled to the system variable $\hat q$ as the input port
with its observable denoted by $\hat F$, which, e.g. 
in optomechnaics, is the force acting on the mechanical oscillator; the
output port with observable $\hat Z$ is the one
projectively measured by
a macroscopic device (e.g. the photodiode)
that produces classical data.

The determining factor behind different measurement tasks
is the quantum noise, arising from quantum fluctuations
in the detector input and output ports. Particularly,
the output-port fluctuation gives rise to the so-called
imprecision noise that quantifies how well
we can probe the system in terms of
measurement precision and rate; the input-port fluctuation
perturbs the system and
leads to the quantum back action noise, which, in the case
of a qubit, can induce dephasing. If the quantum noise
at different times is not correlated, one can
use the master equation approach to study decoherence
of the system or quantum-trajectory approach for analyzing the
system evolution conditional on the measurement outcome
(see, e.g., Ref.\,\cite{Jacobs2006} or
Chapter 4 in Ref.\,\cite{Wiseman2010}).

An alternative approach that can treat
general correlated quantum noise is the linear-response theory
developed by Kubo, and is summarized in his seminal
paper on the fluctuation-dissipation theorem\,\cite{Kubo1966}.
Averin applied it to the dispersive quantum non-demolition
measurement (QND) of a qubit\,\cite{Averin2001, Averin2003},
which has been
further elaborated by Clerk {\it et al.}\,\cite{Clerk2003}
and also extensively reviewed in Ref.\,\cite{Clerk2008}.
Braginsky and Khalili applied this approach to study the
sensitivity of quantum-limited force/displacement sensors\,\cite{Braginsky92}.
In this approach, different dynamical
quantities are related by the susceptibility function
$\chi$ which describes the linear response.
The statistical property of the quantum noise
is quantified by the two-time correlation function, or
equivalently, the frequency-domain noise
spectral density (spectrum)
$\bar S$ for time-invariant (stationary) detectors.

At thermal equilibrium, the spectral density
and susceptibility are connected by the famous
fluctuation-dissipation theorem\,\cite{Callen1951, Kubo1966}.
In contrast, the measurement process considered here
is far away from the thermal equilibrium.
Nevertheless, there is a general Heisenberg-like
uncertainty relation connecting them, which constrains the
imprecision noise and quantum back action noise of the detector, even without knowing
details of the system.
According to Ref.\,\cite{Braginsky92}, such a relation can
be written explicitly as\footnote[3]{there is a minor typo in \cite{Braginsky92}
concerning the sign in front of $\chi_{FF}$.}:
\begin{align}\nonumber
\bar S_{ZZ}(\omega)\bar S_{FF}(\omega)&-|\bar S_{ZF}(\omega)|^2
\ge ({\hbar^2}/{4})
|\chi_{ZF}(\omega)|^2 +\\
&\hbar \big|{\Im}[\bar S_{ZF}^*(\omega)\chi_{ZF}(\omega)-
\chi_{FF}(\omega)\bar S_{ZZ}(\omega)]\big|\,.
\label{eq:inequality}
\end{align}
Here the spectral densities $\bar S_{ZZ}(\omega)$ and
$\bar S_{FF}(\omega)$
quantifies the magnitude of the imprecision noise and
the back action noise at frequency $\omega$, respectively; $\bar S_{ZF}$ quantifies
the cross correlation; the susceptibility $\chi_{ZF}$ describes
the response of the detector output to the system variable;
the input susceptibility $\chi_{FF}$ quantifies the dynamical
back action that modifies the system dynamics, which will be illustrated later using a concrete example;
${\Im}[\cdot]$ means taking the imaginary part.
The mathematical definitions for $\chi$ and $\bar S$
will be given in Eqs.\,\eqref{eq:susceptibility_definition}
and \eqref{eq:spectral_sym}. Note that an uncertainty relation
similar to Eq.\,\eqref{eq:inequality} is
also presented in Ref.\,\cite{Clerk2008};
however, $\chi_{FF}$ has not been
included, and this makes the resulting
uncertainty relation less tight, especially
when $\bar S_{ZF}/\chi_{ZF}$ becomes imaginary.

Applying the above uncertainty relation to measurements of
different systems, one can arrive at some general principles,
regardless of the specific detector used. For example,
in the QND measurement of a qubit,
$\bar S_{ZZ}$ determines the measurement rate $\Gamma_{\rm meas}$
for acquiring information of the qubit state, and $\bar S_{FF}$ sets
the dephasing rate $\Gamma_{\phi}$. Eq.\,\eqref{eq:inequality}
leads to a fundamental quantum limit to the measurement rate: $\Gamma_{\rm meas}\leq \Gamma_{\phi}$---we can measure the qubit at most as fast as we dephase
it\,\cite{Averin2001, Pilgram2002,
Averin2003, Clerk2003, Clerk2008}. In the
classical force/displacement sensing
with a quantum mechanical oscillator,
Eq.\,\eqref{eq:inequality} implies a trade-off
between the imprecision noise and the back-action
noise, which gives rise to the famous Standard Quantum
Limit\,\cite{Braginsky92}. Achieving the quantum limit
requires the detector to be ideal, i.e., having
minimum uncertainty---the thermal excitation of
the detector degrees of freedom and other decoherence effects
become negligible. There are ongoing
experimental efforts towards this goal,
e.g., the most recent results with
optomechanical devices presented
in Refs.\,\cite{Clark2016, Lei2016, Purdy2016, Sudhir2016}.

The main result presented in this paper is to show that
when the detector is at the quantum limit
with minimum uncertainty,
not only do we have Eq.\,\eqref{eq:inequality} attain the equal
sign, but also obtain two new equalities shown in
Eqs.\,\eqref{eq:result1} and \eqref{eq:result2}. By introducing
$\hat z \equiv \hat Z/\chi_{ZF}$ normalized by the output
response, they
can be putted into the following more suggestive form:
\begin{align}\label{eq:result1n}
&\bar S_{zz}(\omega)\bar S_{FF}(\omega)-|\bar S_{zF}(\omega)|^2 = \frac{\hbar^2}{4}\,,\\
&{\Im} [\bar S_{zF}(\omega)] = -{\Im}[\chi_{FF}(\omega)] \bar S_{zz}(\omega) \,.
\label{eq:result2n}
\end{align}
The first equality constraints the
strength of the imprecision noise and the back-action noise, while the second one
relates the cross correlation $\bar S_{zF}$ to the dynamical back action quantified by $\chi_{FF}$. The details will be provided in section \ref{sec:derivation}.

To illustrate implications of Eqs.\,\eqref{eq:result1n}
and \eqref{eq:result2n}, in section \ref{sec:application},
they will be applied to
the typical cavity QED setup, in which a cavity mode is coupled
to either a qubit or a mechanical oscillator. The
key messages are summarized as follows. In the case of
dispersive QND qubit readout, there is an optimal output observable such that $\bar S_{zF}=0$, which leads to
\begin{equation}
 \Gamma_{\rm meas}=\Gamma_{\phi}\,.
\end{equation}
In the case of measuring a mechanical oscillator, the motional
sideband asymmetry reported in several
experiments\,\cite{Safavi-Naeini2012, Weinstein2014, Purdy2015,
Underwood2015, Sudhir2016a} is considered.
One interpretation behind the observed
asymmetry is attributing it to the imprecision-back-action noise correlation
$\bar S_{zF}$\,\cite{Khalili2012, Weinstein2014}. In particular, near the mechanical
resonant frequency $\omega_m$, $\bar S_{zF}$ is purely imaginary and
\begin{equation}\label{eq:S_zF}
 \bar S_{zF}(\omega_m)\approx \pm i\,{\hbar}/{2}\,,
\end{equation}
where $\pm$ depends on the detuning frequency of the laser
with respect to the cavity resonance.
From Eqs.\,\eqref{eq:result1n}
and \eqref{eq:result2n}, such a correlation implies
\begin{equation}\label{eq:chi_FF_S_FF}
{\Im}[\chi_{FF}(\omega_m)] \approx \mp \bar S_{FF}(\omega_m)/\hbar\,.
\end{equation}
Firstly, this indicates that the noise spectra for
the positive and negative
frequencies are highly unbalanced, cf., Eq.\,\eqref{eq:uncertainty_relation1},
which is the case in these experiments. Secondly, since
$\chi_{FF}$ quantifies the dynamical back action
to the mechanical oscillator\,\cite{BuCh2002, Wilson-Rae2007,
Marquardt2007},
the sideband asymmetry observed with the linear measurement
is always accompanied by additional heating or
damping of the mechanical motion.

The outline of this paper goes as follows: in section \ref{sec:definitions},
a brief introduction to the
continuous linear measurement is provided. Additionally,
the formal definitions for the susceptibility function and the
noise spectral density are given; in section \ref{sec:derivation},
the derivation of the general quantum constraints
on detector noise---Eqs.\,\eqref{eq:result1n} and \eqref{eq:result2n} is presented;
in section \ref{sec:application}, these
constraints are illustrated with the examples of quantum
measurements in the cavity QED setup; in section
\ref{sec:discussion}, there will be some discussions about extending the result to more general cases.

\section{Continuous Linear Measurement}\label{sec:definitions}

We now go through the mathematical description
of the continuous linear measurement, and define relevant
quantities, which follows Refs.\,\cite{Clerk2008, Braginsky92, BuCh2002}.
Specifically, the free
Hamiltonian $\hat H_{\rm det}$ of the detector only involves
linear or quadratic functions of canonical coordinates
of which the commutators are classical numbers. The system-detector
interaction $\hat H_{\rm int}$ is in the bilinear form:
\begin{equation}\label{eq:interaction_Hamiltonian}
\hat H_{\rm int}=- \hat q\, \hat F\,.
\end{equation}
Solving the Heisenberg equation of motion leads to the following
solution to the detector observables:
\begin{align}
\label{eq:solution_Z}
\hat Z(t) & =\hat Z^{(0)}(t) + \int_{-\infty}^{+\infty} {\rm d}t' \chi_{ZF}(t-t') \, \hat q(t')\,,\\
\label{eq:solution_F}
\hat F(t) & =\hat F^{(0)}(t)+\int_{-\infty}^{+\infty} {\rm d}t' \chi_{FF}(t-t') \, \hat q(t')\,,
\end{align}
where superscript $(0)$ denotes evolution under the
free Hamiltonian $\hat H_{\rm det}$.
The susceptibility $\chi_{AB}$, quantifying the detector response to
the system variable $\hat q$, is defined as
\begin{equation}\label{eq:susceptibility_definition}
\chi_{AB}(t-t')\equiv (i/\hbar)[\hat A^{(0)}(t),\, \hat B^{(0)}(t')]\Theta(t-t')\,,
\end{equation}
where $\Theta$ is the Heaviside function and
$\hat H_{\rm det}$ is assumed to be time-independent (time-invariant)
so that $\chi_{AB}$ is a function of the time difference $t-t'$.
Note that $\chi_{AB}$ is
not an operator but a classical number, and it only depends on the
free evolution of the detector; both features
are attributable to the detector
being linear. Moving into the frequency domain, we can relate the susceptibility
to the spectral density\,\cite{Clerk2008}:
\begin{equation}\label{eq:sus_spec_relation}
\chi_{AB}(\omega)-\chi^*_{BA}(\omega)=
(i/\hbar)[S_{AB}(\omega)-S_{BA}(-\omega)]\,.
\end{equation}
Here $S_{AB}$ is unsymmetrized spectral density defined through
\begin{equation}\label{eq:spectral_unsym}
{\rm Tr}[\hat \rho_{\rm det}\, \hat A^{(0)}(\omega)
\hat B^{(0)}(\omega')]\equiv 2\pi\,S_{AB}(\omega)
\delta(\omega-\omega')\,,
\end{equation}
where $\hat \rho_{\rm det}$ is the density matrix of
the detector initial state, the Fourier transform
$f(\omega)\equiv \int_{-\infty}^{+\infty} {\rm d}t\,e^{i\omega t} f(t)$,
and ${\rm Tr}[\hat \rho_{\rm det}\hat A^{(0)}]=
{\rm Tr}[\hat \rho_{\rm det}\hat B^{(0)}]=0$ is assumed without loss
of generality.
The symmetrized version of $S_{AB}$, quantifying the fluctuation, is
\begin{equation}\label{eq:spectral_sym}
\bar S_{AB}(\omega) \equiv [S_{AB}(\omega)+S_{BA}(-\omega)]/2\,.
\end{equation}
One special case is when $\hat A$ and $\hat B$ are identical, which
leads to Kubo's formula, using $\hat A=\hat B=\hat F$ as an example:
\begin{equation}\label{eq:Kubo_formula}
\Im[\chi_{FF}(\omega)] =[S_{FF}(\omega)-S_{FF}(-\omega)]/(2\hbar)\,,
\end{equation}
which quantifies the dissipation. In the thermal equilibrium,
$S_{FF}(\omega)$ and $S_{FF}(-\omega)$
differ from each other by the Boltzmann factor $e^{\hbar \omega/(k_B T)}$
with $T$ being temperature, and
$\bar S_{FF}(\omega)$ is thus related to $\Im[\chi_{FF}(\omega)]$
by the fluctuation-dissipation theorem. For the measurement process
far from thermal equilibrium, we generally have
\begin{equation}\label{eq:uncertainty_relation1}
\bar S_{FF}(\omega)\ge
\hbar |\Im[\chi_{FF}(\omega)]|\,,
\end{equation}
in which the equal sign is achieved
when either $S_{FF}(\omega)$ or $S_{FF}(-\omega)$ vanishes\,\cite{Braginsky92}.
This relation constraints the quantum fluctuation in either $\hat Z$ or $\hat F$
individually; the uncertainty relation Eq.\,\eqref{eq:inequality}
connects both together.

\section{The General Quantum Constraints}\label{sec:derivation}

After defining key quantities,
we come to the derivation of Eq.\,\eqref{eq:inequality}
and the main result Eqs.\,\eqref{eq:result1n} and \eqref{eq:result2n}---the general quantum constraints
on detector noise.
It follows the standard approach outlined in Ref.\,\cite{Braginsky92}, but uses
unsymmetrized spectral density as the starting point, which allows directly showing
the condition for achieving the quantum limit. Define the following auxiliary
operator:
\begin{equation}\label{eq:Q}
\hat Q\equiv \int_{-\infty}^{+\infty} {\rm d}\omega [\alpha^*(\omega) \hat Z^{(0)}(\omega)+\beta^*(\omega)\hat F^{(0)}(\omega)]\,,
\end{equation}
where $\alpha, \beta$ are some functions.
The norm of $\hat Q$ is positive definite, i.e.,
$||\hat Q||^2\equiv {\rm Tr}[\hat \rho_{\rm det}
\hat Q \,\hat Q^{\dag}]\ge 0$,
which, in terms of unsymmetrized spectral density,
reads
\begin{equation}\label{eq:norm_Q_spec}
\int_{-\infty}^{+\infty}{\rm d}\omega\, [\alpha^*,\, \beta^*]
\left[\begin{array}{cc}
        S_{ZZ}(\omega) & S_{ZF}(\omega) \\
        S_{ZF}^*(\omega) & S_{FF}(\omega)
      \end{array}
\right]\left[\begin{array}{c}
               \alpha \\
               \beta
             \end{array}
\right]\ge 0\,.
\end{equation}
It needs to be satisfied for arbitrary
$\alpha$ and $\beta$, which implies
\begin{equation}\label{eq:determinant}
S_{ZZ}(\omega)S_{FF}(\omega) - |S_{ZF}(\omega)|^2\ge 0\,.
\end{equation}
Using Eqs.\,\eqref{eq:sus_spec_relation} and
\eqref{eq:spectral_sym}, we can rewrite it as
\begin{align}\nonumber
\{\bar S_{ZZ}(\omega)\pm \hbar \Im&[\chi_{ZZ}(\omega)]\}\{\bar S_{FF}(\omega)\pm \hbar\Im[\chi_{FF}(\omega)]\} \ge \\
&\big|\bar S_{ZF}(\omega)\pm\frac{\hbar}{2i} [\chi_{ZF}(\omega)-\chi_{FZ}^*(\omega)]\big|^2\,.
\label{eq:determinant2}
\end{align}
Here $\pm$ comes from that Eq.\,\eqref{eq:determinant} needs
to be satisfied for both positive and negative frequencies.

In order for $\hat F$ and $\hat Z$ to be the input and output
observables of the detector, the susceptibilities cannot take arbitrary value.
Because the macroscopic device, illustrated in Fig.\,\ref{fig:measurement_model},
needs to make projective measurement of $\hat Z$ continuously in time, which produces
a classical data stream. This means the final $\hat Z$ after
interacting with the system, shown in Eq.\,\eqref{eq:solution_Z},
can be precisely measured at different times, which happens only if
\begin{equation}\label{eq:measurability}
[\hat Z(t),\,\hat Z(t')] =0\quad \forall\,t,\,t'.
\end{equation}
In Ref.\,\cite{BuCh2002}, the authors called this as the condition of
simultaneous measurability, and further showed that it implies
$[\hat Z^{(0)}(t),\,\hat Z^{(0)}(t')] = [\hat F^{(0)}(t),\,\hat Z^{(0)}(t')] \Theta(t-t')=0$, i.e.,
\begin{equation}\label{eq:chi_zz_chi_fz}
\chi_{ZZ}(\omega)=\chi_{FZ}(\omega)=0\,.
\end{equation}
Taking this into account, Eq.\,\eqref{eq:determinant2} leads to
\begin{align}\nonumber
\bar S_{ZZ}(\omega)\bar S_{FF}(\omega)&-|\bar S_{ZF}(\omega)|^2
\ge \frac{\hbar^2}{4}
|\chi_{ZF}(\omega)|^2 \pm\\
&\hbar {\Im}[\bar S_{ZF}^*(\omega)\chi_{ZF}(\omega)-
\chi_{FF}(\omega)\bar S_{ZZ}(\omega)]\,.
\label{eq:inequality_pm}
\end{align}
Since the inequality has to be valid for both plus sign and
minus sign in front of $\hbar\Im[\cdot]$, it becomes equivalent to
the uncertainty relation Eq.\,\eqref{eq:inequality}.

When the detector is
at the quantum limit, Eq.\,\eqref{eq:determinant} takes the minimum, i.e.,
\begin{equation}\label{eq:determinant_equality}
\left\{S_{ZZ}(\omega)S_{FF}(\omega) - |S_{ZF}
(\omega)|^2\right\}_{\rm quantum\,limit}=0\,.
\end{equation}
Equivalently, this gives rise to two equalities for
either plus sign or minus sign in
Eq.\,\eqref{eq:inequality_pm}.
Taking their sum and difference, we obtain
\begin{align}\label{eq:result1}
&\bar S_{ZZ}(\omega)\bar S_{FF}(\omega)-|\bar S_{ZF}(\omega)|^2 = \frac{\hbar^2}{4}
|\chi_{ZF}(\omega)|^2\,,\\
&\Im[\bar S_{ZF}^*(\omega)\chi_{ZF}(\omega)-\chi_{FF}(\omega)\bar S_{ZZ}(\omega)]=0\,,
\label{eq:result2}
\end{align}
which are reduced to Eqs.\,\eqref{eq:result1n} and \,\eqref{eq:result2n}
after normalizing $\hat Z$ by the susceptibility $\chi_{ZF}$.

One can show explicitly that the quantum limit is achieved when the detector is in the pure, stationary, Gaussian state---the multi-mode squeezed
state (see, e.g., Ref.\,\cite{Blow1990}):
\begin{equation}\label{eq:sqeezed_state}
|\xi\rangle \equiv
\exp\left\{\int_0^{+\infty} \frac{{\rm d}\omega}{2\pi} [\xi(\omega) \hat d^{\dag}(\omega)\hat d^{\dag}(-\omega)-{\rm h.c.}]\right\}|0\rangle\,.
\end{equation}
Here we have used the fact that the detector is linear with its canonical coordinates having classical-number commutators, and
thus it can be modelled as a collection of bosonic modes (bosonic field); $\hat d(\omega)$ is the corresponding 
annihilation operator of the detector
mode at frequency $\omega$; $\xi(\omega)$ describes the squeezing factor at different frequencies;
h.c. denotes Hermitian conjugate; $|0\rangle$ is the vacuum state.
In quantum optics, the zero frequency in Eq.\,\eqref{eq:sqeezed_state} coincides
with one half of the pump frequency
of the optical parametric oscillator that produces the squeezed state.

Using the fact that $\hat Z^{(0)}(t)$ and $\hat F^{(0)}(t)$ are Hermitian,
we can rewrite their Fourier transform in terms of $\hat d(\omega)$ and $\hat d^{\dag}(-\omega)$:
\begin{align}\label{eq:Z_F_d}
\hat Z^{(0)}(\omega)&={\cal Z}(\omega) \hat d(\omega)+{\cal Z}^*(-\omega)\hat d^{\dag}(-\omega)\,,\\
\hat F^{(0)}(\omega)&={\cal F}(\omega) \hat d(\omega)+{\cal F}^*(-\omega)\hat d^{\dag}(-\omega)\,,
\end{align}
with some coefficients ${\cal Z}$ and ${\cal F}$. One can then find that
\begin{equation}\label{eq:determinat3}
\det  \left[\begin{array}{cc}
                    \langle \xi|\hat Z(\omega)\hat Z^{\dag}(\omega')|\xi\rangle & \langle \xi|\hat Z(\omega)\hat F^{\dag}(\omega')|\xi\rangle \\
                    \langle \xi|\hat F(\omega)\hat Z^{\dag}(\omega')|\xi\rangle & \langle \xi|\hat F(\omega)\hat F^{\dag}(\omega')|\xi\rangle
                  \end{array}\right]
=0\,,
\end{equation}
where the superscript (0) is omitted. This is equivalent to Eq.\,\eqref{eq:determinant_equality} according
to the definition of the spectral density shown in Eq.\,\eqref{eq:spectral_unsym}.

\section{Application to cavity QED}\label{sec:application}

\begin{figure}[!b]
\includegraphics[width=\columnwidth]{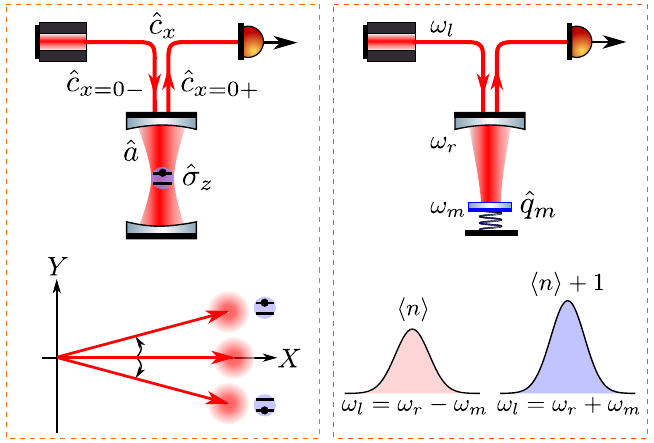}
\caption{Schematics for the dispersive qubit readout (left) in which the state-dependent
phase shift is inferred by probing the phase quadrature $\hat Y$ of the cavity mode, and the
sideband asymmetry measurement (right) in which we tune the laser frequency to
selectively measure the Stokes and anti-Stokes sidebands scattered by
the mechanical motion. In both cases, the cavity mode $\hat a$
is continuously driven by the external field $\hat c_{x}$, which contains both the
coherent amplitude and quantum fluctuation. We use $\hat c_{x=0-}$ ($\hat c_{x=0+}$)
to denote the ingoing (outgoing) field right before (after) interacting with
the cavity mode. The outgoing part is monitored by a photodiode using homodyne
detection.
\label{fig:configuration}}
\end{figure}

To illustrate the above result, let us look at the cavity QED setup shown schematically in Fig.\,\ref{fig:configuration}. The system can either be a qubit or
a mechanical oscillator. The detector consists of a single cavity mode
$\hat a$ and external continuum field $\hat c_x$
with the central frequency defined by the laser frequency $\omega_l$ which
can be detuned
from the cavity resonant frequency $\omega_r$. Its Hamiltonian in the
rotating frame of the laser frequency is given by (see, e.g., section 2 in \cite{Yanbei:Review})
\begin{align}\nonumber
\hat H_{\rm det}=\hbar (\omega_r -\omega_l) \hat a^{\dag}\hat a  &- i \hbar \int_{-\infty}^{+\infty}{\rm d}x \,\hat c_x^{\dag} \frac{\partial \hat c_x}{\partial x} \\ & +
i\hbar \sqrt{2\gamma}\,(\hat a^{\dag} \hat c_{x=0} -  \hat a
\,\hat c_{x=0}^{\dag})\,.
\label{eq:H_det}
\end{align}
Here subscript $x$ in $\hat c_x$ labels the field degree of freedom at different locations; $\hat c_{x=0}$ denotes the one directly coupled to the cavity mode at a rate $\gamma$. This is the
same Hamiltonian for a one-sided cavity in the standard input-output (has a different meaning
from the one used here) formalism\,\cite{Gardiner1985}.

The interaction Hamiltonian in two cases, with the system to be a qubit (dispersive-coupling regime\,\cite{Blais2004, Wallraff2005}) and a mechanical oscillator (optomechanical coupling\,\cite{Yanbei:Review, Aspelmeyer2014}), are
\begin{equation}\label{eq:H_int}
\hat H^{\rm qubit}_{\rm int}=-\hbar \frac{g_0^2}{\omega_l-\omega_{01}} \hat \sigma_z \hat a^{\dag}\hat a\,,\quad \hat H^{\rm mech}_{\rm int}=-\hbar \frac{\omega_r}{L} \hat q_m \hat a^{\dag}\hat a\,,
\end{equation}
where $g_0$ is the cavity-qubit coupling rate in the Jaynes-Cummings model,
$\omega_{01}$ is the transition frequency between two energy levels,
$\hat \sigma_z$ is the Pauli operator, $L$ is the cavity length, and $\hat q_m$ is the position
of the mechanical oscillator.

For both cases, the input observable $\hat F$
is proportional to the cavity photon number $\hat n_{\rm cav}=\hat a^{\dag}\hat a$ by comparing with Eq.\,\eqref{eq:interaction_Hamiltonian}, and thus in general we can write
$\hat F = \hbar g \hat n_{\rm cav}$ with $g$ depending on the specific system.
Due to pumping from the laser, the mean cavity photon
number $\bar n_{\rm cav}$ is much larger than one. The standard approach is linearizing
$\hat n_{\rm cav}$ and keeping the perturbed part that is proportional to the
amplitude quadrature $\hat X\equiv (\hat a+\hat a^{\dag})/\sqrt{2}$. The resulting
linearized $\hat F$ reads
\begin{equation}\label{eq:F_QED}
\hat F =\hbar g \sqrt{2\bar n_{\rm cav}}\,\hat X \equiv \hbar \bar g \hat X\,.
\end{equation}
Additionally, given homodyne detection of the outgoing field $\hat c_{x=0+}$,
the output observable $\hat Z$ can be written as
\begin{equation}\label{eq:Z_QED}
\hat Z = \cos\theta\, \hat X_{\rm out} +\sin\theta\, \hat Y_{\rm out}\,,
\end{equation}
where $\theta$ depends on the phase of the local oscillator in the homodyne detection,
$\hat X_{\rm out}\equiv (\hat c_{x=0+}+\hat c_{x=0+}^{\dag})/\sqrt{2}$
and $\hat Y_{\rm out}\equiv (\hat c_{x=0+}-\hat c_{x=0+}^{\dag})/(\sqrt{2}i)$
are the amplitude quadrature and phase quadrature of the outgoing field.

To derive the susceptibilities and spectral densities,
cf., Eqs.\,\eqref{eq:susceptibility_definition} and \eqref{eq:spectral_unsym},
we only need to solve the Heisenberg equation for the cavity mode and external
field under the free evolution of $\hat H_{\rm det}$, according to Ref.\,\cite{Gardiner1985}:
\begin{align}\label{eq:EOMs_det}
\dot {\hat a}(t) & =-(\gamma-i\Delta)\hat a(t)+\sqrt{2\gamma}\, \hat c_{x=0-}(t)\,,\\
\hat c_{x=0+}(t) &=\hat c_{x=0-}(t)-\sqrt{2\gamma}\,\hat a(t)\,,
\end{align}
where $\Delta\equiv \omega_l -\omega_r$ is
the laser detuning frequency with respect to
the cavity resonance. Solving these equations in the frequency domain,
we can represent the cavity mode $\hat a$ and the outgoing field
$\hat c_{x=0+}$ in terms of the ingoing field $\hat c_{x=0-}$.

Without using the non-classical squeezed light, the spectral density for $\hat c_{x=0-}$
is simply $S_{\hat c\hat c^{\dag}}(\omega)=1$ (vacuum fluctuation)
and $S_{\hat c^{\dag}\hat c}(\omega)=0$, from which we obtain the
relevant spectra (double-sided):
\begin{align}\label{eq:spec_optomech}
\bar S_{ZZ}(\omega)&=\frac{1}{2},\\
\bar S_{ZF}(\omega)&=\frac{\hbar \bar g\sqrt{\gamma}[\Delta\sin\theta- (i\omega-\gamma) \cos\theta] }{(\omega-\Delta+i\gamma)(\omega+\Delta+i\gamma)}\label{eq:spec_SZF}\,,\\
\bar S_{FF}(\omega)&=\frac{2\hbar^2 \bar g^2 \gamma(\gamma^2+\Delta^2+\omega^2)}{[(\omega-\Delta)^2+\gamma^2][(\omega+\Delta)^2+\gamma^2]}\,,
\end{align}
and the susceptibilities:
\begin{align}\label{eq:sus_chiZF}
\chi_{ZF}(\omega)&=-\frac{2\bar g\sqrt{\gamma}[\Delta\cos\theta+ (i\omega-\gamma) \sin\theta] }{(\omega-\Delta+i\gamma)(\omega+\Delta+i\gamma)}\,,\\
\chi_{FF}(\omega)&=\frac{2\hbar \bar g^2\Delta }{(\omega-\Delta+i\gamma)(\omega+\Delta+i\gamma)}\,.
\end{align}
One can check that they indeed satisfy the general quantum
constraints Eqs.\,\eqref{eq:result1n} and \eqref{eq:result2n}.

For the qubit readout, by introducing $\hat z\equiv \hat Z/\chi_{ZF}$, we have
\begin{equation}
\hat z(\omega)=\hat z^{(0)}(\omega)+\hat \sigma_z \delta(\omega)\,.
\end{equation}
Here it uses the fact that $\hat \sigma_z$ is a QND observable
and remains constant in time. Therefore, the output responds to
the signal only near DC with $\omega\approx 0$. For measurement with a
finite duration, the delta function $\delta(\omega)$ is
approximately equal to the integration time. The measurement
rate is defined by the inverse of the integration time that is required
to reach signal-to-noise ratio equal to one half,
using the convention in Ref.\,\cite{Clerk2008}:
\begin{equation}
 \Gamma_{\rm meas}\equiv {1}/[{2\,\bar S_{zz}(0)}]\,.
\end{equation}
The fluctuation in the cavity photon number induces a random AC Stark
shift on the energy level, which causes dephasing,
cf., Eq.\,\eqref{eq:H_int}. With measurement
much longer than the cavity storage time, only the low-frequency
part of the back action noise spectrum is relevant, and according to
Ref.\,\cite{Clerk2008}, the dephasing rate is
\begin{equation}
 \Gamma_{\phi}\equiv ({2}/{\hbar^2}) \bar S_{FF}(0)\,.
\end{equation}
At the quantum limit, the ratio between these two rates is given by, cf., Eqs.\,\eqref{eq:result1n}, \eqref{eq:spec_SZF},
and \eqref{eq:sus_chiZF},
\begin{equation}
 \frac{\Gamma_{\phi}}{\Gamma_{\rm meas}}=1+\frac{4\bar S_{zF}^2(0)}{\hbar^2}=1+\left(\frac{\Delta\sin\theta+\gamma\cos\theta}
 {\Delta\cos\theta-\gamma\sin\theta}\right)^2\,.
\end{equation}
The optimal readout quadrature for reaching
$\Gamma_{\phi}=\Gamma_{\rm meas}$ is therefore the one satisfying
\begin{equation}
 \theta_{\rm opt}=-\arctan(\gamma/\Delta)\,.
\end{equation}
When the cavity is tuned with $\Delta=0$, $\theta_{\rm opt}=\pm\pi/2$ and
the output phase quadrature
is the optimal one, cf., Eq.\,\eqref{eq:Z_QED}, while for a
large detuning $\Delta \gg \gamma$, $\theta_{\rm opt}\approx 0$
and we need to measure the output
amplitude quadrature. This result can be generalized
to more complicated measurement setups with, e.g., multiple coupled cavities.
Because such a measurement is near DC and $\bar S_{zF}(0)$ is real,
we can always find the right readout quadrature such that $\bar S_{zF}(0)=0$, which makes $\Gamma_{\rm meas}=\Gamma_{\phi}$ at the quantum limit.

We now switch
to the case of measuring mechanical motion. In contrast to the qubit readout,
the position $\hat q_m$ of the mechanical oscillator
is not a QND observable, and the back-action noise will appear in the output:
\begin{align}\nonumber
 \hat z(\omega) & = \hat z^{(0)}(\omega)+\hat q_m(\omega)\\
 & = \hat z^{(0)}(\omega)+\chi_{qq}(\omega) [\hat F^{(0)}(\omega)+\hat F_{\rm th}(\omega)]\,,
\end{align}
where it uses the fact that $\hat q_m = \chi_{qq}[\hat F^{(0)}
+\hat F_{\rm th}]$ with $\hat F_{\rm th}$ being the thermal Langevin force and $\chi_{qq}$ being the mechanical susceptibility modified by the detector input susceptibility:
\begin{equation}
\chi_{qq}(\omega)=\chi_{qq}^{(0)}(\omega)/[1-\chi_{qq}^{(0)}(\omega)\chi_{FF}(\omega)]\,,
\end{equation}
in which $\chi_{qq}^{(0)}$ is the original (bare) mechanical
susceptibility.
Notice that $\chi_{FF}$ is often referred to as
the optical spring coefficient or dynamical back action
in the literature, which introduces additional
heating or damping to the mechanical motion. This has been utilized in the optomechanical sideband cooling experiments\,\cite{Marquardt2007, Wilson-Rae2007, Yanbei:Review, Aspelmeyer2014}.

The total
output spectrum reads
\begin{equation}\label{eq:spec_output}
 \bar S_{zz}^{\rm tot}(\omega)=\bar S_{zz}(\omega)+ 2 \Re[\chi^*_{qq}(\omega)\bar S_{zF}(\omega)] + \bar S_{qq}(\omega)\,,
\end{equation}
where $\Re[\cdot]$ means taking the real part.
According to Refs.\,\cite{Khalili2012, Weinstein2014},
it is the second term, i.e.,
the cross correlation between the imprecision
noise and the back action noise, gives rise to the observed asymmetry. From
Eqs.\,\eqref{eq:spec_SZF} and \eqref{eq:sus_chiZF}, we have
\begin{equation}
 \bar S_{zF}(\omega)=-\frac{\hbar}{2}\left[\frac{\Delta\sin\theta- (i\omega-\gamma) \cos\theta}{\Delta\cos\theta+ (i\omega-\gamma) \sin\theta}\right]\,.
\end{equation}
Those experiments reported in Refs.\,\cite{Safavi-Naeini2012, Weinstein2014, Purdy2015, Underwood2015, Sudhir2016a} are operating in the resolved sideband regime with the cavity bandwidth much smaller than the mechanical resonant frequency, i.e.,
$\gamma\ll\omega_m$, and also
 the detuning frequency $\Delta=\pm \omega_m$. Therefore, near
the mechanical resonant frequency $\omega_m$,
$\bar S_{zF}$ is
approximately equal to the one shown in Eq.\,\eqref{eq:S_zF},
which leads to
\begin{equation}
 2\Re[\chi^*_{qq}(\omega_m)\bar S_{zF}(\omega_m)]\big|_{\Delta=\pm \omega_m}\approx \pm \hbar \Im[\chi_{qq}(\omega_m)]\,.
\end{equation}
Since $\bar S_{qq}(\omega_m)=\hbar(2\langle n \rangle +1)\Im[\chi_{qq}(\omega_m)]$
with the mean occupation number $\langle n\rangle = 1/(e^{\hbar\omega_m/k_B T}-1)$
from the fluctuation-dissipation theorem, the above term either doubles or cancels
the contribution from the zero-point fluctuation of the mechanical oscillator,
which induces the
asymmetry. According to Eq.\,\eqref{eq:result2n}, such an imaginary
correlation
is always associated with the dynamical back action quantified by $\chi_{FF}$:
\begin{equation}\label{eq:im_chi_FF}
\Im[\chi_{FF}(\omega_m)]_{\Delta=\pm \omega_m}\approx \mp {\hbar \bar g^2}/{\gamma}\approx \mp {\bar S_{FF}(\omega_m)}/{\hbar}\,,
\end{equation}
as mentioned earlier in
Eq.\,\eqref{eq:chi_FF_S_FF}.

Before leaving this example, there is one comment motivated by
Ref.\,\cite{Buchmann2016}.
The imaginary cross correlation is only detectable near the mechanical resonance
when using the homodyne readout, because $\chi_{qq}(\omega_m)$
is also imaginary, which makes the second term in Eq.\eqref{eq:spec_output} nonzero.
If measuring far away from the resonance or
the oscillator were lossless, we will need to apply the
synodyne readout scheme
presented in Ref.\,\cite{Buchmann2016} to probe such a quantum correlation.

\section{Discussion}\label{sec:discussion}

The above discussion showed the general quantum constraints for the detector noise in linear continuous measurement. Particularly, the noise spectral densities (quantifying the quantum fluctuation) and susceptibilities (quantifying the linear response/dissipation) of detectors at the quantum limit were shown to be related by two equalities, which can be viewed as a generalization of the fluctuation-dissipation theorem to the non-equilibrium quantum measurement processes. The result is general and can be applied to different measurement setups, and we have seen two examples in cavity QED.

One last point worthy mentioning is that so far we have only covered linear detectors with single input and single output, which is the case for most experiments mentioned earlier.
The result can be generalized to
multiple-input-multiple-output (MIMO)
detectors through the following identity, which is a generalization
of Eq.\,\eqref{eq:determinat3},
\begin{equation}\label{eq:MIMO}
\det \langle \xi | \hat {\bf A}(\omega)
\hat {\bf A}^{\dag}(\omega')|\xi \rangle =0\,,
\end{equation}
where $\hat {\bf A}^{\dag}= (\hat Z^{(0)}_1, \hat F_1^{(0)}, \cdots,\hat Z^{(0)}_{N}, \hat F_{N}^{(0)})^{\dag}$ with $N$ being the number of ports,
and the condition of simultaneous measurability:
\begin{equation}
[\hat Z_k(t),\,\hat Z_l(t')]=0\quad \forall\, t,\, t'
\end{equation}
with $k, l=1, \cdots, N$. This follows the same logic
as deriving the uncertainty relation for MIMO detectors and obtaining Eq.\,\eqref{eq:inequality} as one special case in Ref.\,\cite{Braginsky92}.

\section{Acknowledgement}

H.M. would like to thank Farid Khalili and Yanbei Chen for important comments and discussions. This research is supported by EU Marie-Curie Fellowship and UK STFC Ernest Rutherford Fellowship.


\bibliography{references}


\end{document}